%% file: main.tex
\documentclass[twocolumn,
              superscriptaddress,
              prc,
              showkeys,
              showpacs,
              nofootinbib,
              notitlepage,
              floatfix,
              preprintnumbers,
              ]{revtex4-2}

\input{packages}
\input{aliases}

\begin{document}

\preprint{Prepared for submission to PRC}

\title{Spectral Measurement of the $^{214}$Bi \bdec\ \\ to the $^{214}$Po Ground State with XENONnT}
\include{authors}

\date{\today}

\input{values}

\begin{abstract}
We report the measurement of the \Bi\ beta-decay spectrum to the ground state of \Po\ using the XENONnT detector. This decay is classified as first-forbidden non-unique, for which theoretical predictions require detailed nuclear structure modeling. A dedicated identification algorithm isolates a high-purity sample of ground-state beta-decays, explicitly excluding events with associated gamma-rays emission. By comparing the measured spectrum, which covers energies up to \SI{\get{BetaEndPoint}}{\MeV}, with several nuclear models, we find that the prediction based on the conserved vector current (CVC) hypothesis provides the best description of the data. Using this dataset, we additionally derive charge and light yield curves for electronic recoils, extending detector response modeling up to the MeV scale.

\end{abstract}

\maketitle

\input{sections/1_Introduction}
\input{sections/2_Theory}

\input{sections/3_Experimental}

\input{sections/4_Results}
\input{sections/5_Conclusion}

\begin{acknowledgments}
We thank X. Mougeot, M. Ramalho, and J. Suhonen for kindly sharing data and for valuable discussions.

We thank ISOLDE team at CERN for assisting in carrying out the implantation for the \Rn\ source.

We gratefully acknowledge support from the National Science Foundation, Swiss National Science Foundation, German Ministry for Education and Research, Max Planck Gesellschaft, Deutsche Forschungsgemeinschaft, Helmholtz Association, Dutch Research Council (NWO), Fundacao para a Ciencia e Tecnologia, Weizmann Institute of Science, Binational Science Foundation, Région des Pays de la Loire, Knut and Alice Wallenberg Foundation, Kavli Foundation, JSPS Kakenhi, JST FOREST Program, and ERAN in Japan, Tsinghua University Initiative Scientific Research Program, DIM-ACAV+ Région Ile-de-France, and Istituto Nazionale di Fisica Nucleare. This project has received funding/support from the European Union’s Horizon 2020 research and innovation program under the Marie Skłodowska-Curie grant agreement No 860881-HIDDeN.

We gratefully acknowledge support for providing computing and data-processing resources of the Open Science Pool and the European Grid Initiative, at the following computing centers: the CNRS/IN2P3 (Lyon - France), the Dutch national e-infrastructure with the support of SURF Cooperative, the Nikhef Data-Processing Facility (Amsterdam - Netherlands), the INFN-CNAF (Bologna - Italy), the San Diego Supercomputer Center (San Diego - USA) and the Enrico Fermi Institute (Chicago - USA). We acknowledge the support of the Research Computing Center (RCC) at The University of Chicago for providing computing resources for data analysis.

We thank the INFN Laboratori Nazionali del Gran Sasso for hosting and supporting the XENON project.
\end{acknowledgments}

\bibliography{biblio}
\input{sections/9_Appendix}

\end{document}

%% file: packages.tex
\usepackage{graphicx}
\usepackage{dcolumn}
\usepackage{bm}
\usepackage{lineno}
\usepackage[perpage]{footmisc}
\usepackage{graphicx}
\usepackage[colorlinks = true,
            linkcolor = purple,
            urlcolor  = blue,
            citecolor = purple,
            anchorcolor = blue]{hyperref}
\usepackage[dvipsnames]{xcolor}
\usepackage{amssymb}
\usepackage[nointegrals]{wasysym}
\usepackage{amsthm}
\usepackage{amsmath}
\usepackage{textcomp}
\usepackage{mathtools}
\usepackage{multirow}
\usepackage{upgreek}
\usepackage{isotope}
\usepackage{float}
\usepackage{soul, ulem}

\usepackage[separate-uncertainty,retain-explicit-plus,per-mode = symbol]{siunitx}
\usepackage{url}
\usepackage{array}
\usepackage{nicefrac}
\usepackage{lipsum}  
\usepackage{adjustbox}
\usepackage{comment}
\usepackage{environ}

\usepackage{booktabs}
\usepackage{enumitem}
\setlist{nosep}
\usepackage{orcidlink}
\usepackage{currfile}

%% file: aliases.tex
\newcommand{\Ra}{\isotope[226]{Ra}}
\newcommand{\Rn}{\isotope[222]{Rn}}
\newcommand{\Bi}{\isotope[214]{Bi}}
\newcommand{\Po}{\isotope[214]{Po}}
\newcommand{\PbTwoOneZero}{\isotope[210]{Pb}}
\newcommand{\BiPo}{\isotope[214]{BiPo}}
\newcommand{\Pb}{\isotope[214]{Pb}}
\newcommand{\Ar}{\isotope[37]{Ar}}

\newcommand{\KrEightFive}{\isotope[85]{Kr}}

\newcommand{\adec}{alpha-decay}
\newcommand{\bdec}{beta-decay}

\newcommand{\adecs}{alpha-decays}
\newcommand{\bdecs}{beta-decays}
\newcommand{\gdecs}{gamma-decays}

\newcommand{\Bdec}{Beta-decay}

%% file: authors.tex

\newcommand{\bologna}{\affiliation{Department of Physics and Astronomy, University of Bologna and INFN-Bologna, 40126 Bologna, Italy}}
\newcommand{\chicago}{\affiliation{Department of Physics, Enrico Fermi Institute \& Kavli Institute for Cosmological Physics, University of Chicago, Chicago, IL 60637, USA}}
\newcommand{\coimbra}{\affiliation{LIBPhys, Department of Physics, University of Coimbra, 3004-516 Coimbra, Portugal}}
\newcommand{\columbia}{\affiliation{Physics Department, Columbia University, New York, NY 10027, USA}}
\newcommand{\lngs}{\affiliation{INFN-Laboratori Nazionali del Gran Sasso and Gran Sasso Science Institute, 67100 L'Aquila, Italy}}
\newcommand{\mainz}{\affiliation{Institut f\"ur Physik \& Exzellenzcluster PRISMA$^{+}$, Johannes Gutenberg-Universit\"at Mainz, 55099 Mainz, Germany}}
\newcommand{\mpik}{\affiliation{Max-Planck-Institut f\"ur Kernphysik, 69117 Heidelberg, Germany}}
\newcommand{\munster}{\affiliation{Institut f\"ur Kernphysik, University of M\"unster, 48149 M\"unster, Germany}}
\newcommand{\nikhef}{\affiliation{Nikhef and the University of Amsterdam, Science Park, 1098XG Amsterdam, Netherlands}}
\newcommand{\nyuad}{\affiliation{New York University Abu Dhabi - Center for Astro, Particle and Planetary Physics, Abu Dhabi, United Arab Emirates}}
\newcommand{\purdue}{\affiliation{Department of Physics and Astronomy, Purdue University, West Lafayette, IN 47907, USA}}
\newcommand{\rice}{\affiliation{Department of Physics and Astronomy, Rice University, Houston, TX 77005, USA}}
\newcommand{\stockholm}{\affiliation{Oskar Klein Centre, Department of Physics, Stockholm University, AlbaNova, Stockholm SE-10691, Sweden}}
\newcommand{\subatech}{\affiliation{SUBATECH, IMT Atlantique, CNRS/IN2P3, Nantes Universit\'e, Nantes 44307, France}}
\newcommand{\torino}{\affiliation{INAF-Astrophysical Observatory of Torino, Department of Physics, University  of  Torino and  INFN-Torino,  10125  Torino,  Italy}}
\newcommand{\ucsd}{\affiliation{Department of Physics, University of California San Diego, La Jolla, CA 92093, USA}}
\newcommand{\wis}{\affiliation{Department of Particle Physics and Astrophysics, Weizmann Institute of Science, Rehovot 7610001, Israel}}
\newcommand{\zurich}{\affiliation{Physik-Institut, University of Z\"urich, 8057  Z\"urich, Switzerland}}
\newcommand{\paris}{\affiliation{LPNHE, Sorbonne Universit\'{e}, CNRS/IN2P3, 75005 Paris, France}}
\newcommand{\freiburg}{\affiliation{Physikalisches Institut, Universit\"at Freiburg, 79104 Freiburg, Germany}}
\newcommand{\napels}{\affiliation{Department of Physics ``Ettore Pancini'', University of Napoli and INFN-Napoli, 80126 Napoli, Italy}}
\newcommand{\nagoya}{\affiliation{Kobayashi-Maskawa Institute for the Origin of Particles and the Universe, and Institute for Space-Earth Environmental Research, Nagoya University, Furo-cho, Chikusa-ku, Nagoya, Aichi 464-8602, Japan}}
\newcommand{\laquila}{\affiliation{Department of Physics and Chemistry, University of L'Aquila, 67100 L'Aquila, Italy}}
\newcommand{\tokyo}{\affiliation{Kamioka Observatory, Institute for Cosmic Ray Research, and Kavli Institute for the Physics and Mathematics of the Universe (WPI), University of Tokyo, Higashi-Mozumi, Kamioka, Hida, Gifu 506-1205, Japan}}
\newcommand{\kobe}{\affiliation{Department of Physics, Kobe University, Kobe, Hyogo 657-8501, Japan}}
\newcommand{\kit}{\affiliation{Institute for Astroparticle Physics, Karlsruhe Institute of Technology, 76021 Karlsruhe, Germany}}
\newcommand{\tsinghua}{\affiliation{Department of Physics \& Center for High Energy Physics, Tsinghua University, Beijing 100084, P.R. China}}
\newcommand{\ferrara}{\affiliation{INFN-Ferrara and Dip. di Fisica e Scienze della Terra, Universit\`a di Ferrara, 44122 Ferrara, Italy}}
\newcommand{\groningen}{\affiliation{Nikhef and the University of Groningen, Van Swinderen Institute, 9747AG Groningen, Netherlands}}
\newcommand{\westlake}{\affiliation{Department of Physics, School of Science, Westlake University, Hangzhou 310030, P.R. China}}
\newcommand{\shenzhen}{\affiliation{School of Science and Engineering, The Chinese University of Hong Kong (Shenzhen), Shenzhen, Guangdong, 518172, P.R. China}}
\newcommand{\coimbrapoli}{\affiliation{Coimbra Polytechnic - ISEC, 3030-199 Coimbra, Portugal}}
\newcommand{\uniheidelberg}{\affiliation{Physikalisches Institut, Universit\"at Heidelberg, Heidelberg, Germany}}
\newcommand{\roma}{\affiliation{INFN-Roma Tre, 00146 Roma, Italy}}
\newcommand{\bucknell}{\affiliation{Department of Physics \& Astronomy, Bucknell University, Lewisburg, PA, USA}}

\author{E.~Aprile\,\orcidlink{0000-0001-6595-7098}}\columbia
\author{J.~Aalbers\,\orcidlink{0000-0003-0030-0030}}\groningen
\author{K.~Abe\,\orcidlink{0009-0000-9620-788X}}\tokyo
\author{M.~Adrover\,\orcidlink{0123-4567-8901-2345}}\zurich
\author{S.~Ahmed Maouloud\,\orcidlink{0000-0002-0844-4576}}\paris
\author{L.~Althueser\,\orcidlink{0000-0002-5468-4298}}\munster
\author{B.~Andrieu\,\orcidlink{0009-0002-6485-4163}}\paris
\author{E.~Angelino\,\orcidlink{0000-0002-6695-4355}}\torino\lngs
\author{D.~Ant\'on~Martin\,\orcidlink{0000-0001-7725-5552}}\chicago
\author{S.~R.~Armbruster\,\orcidlink{0009-0009-6440-1210}}\mpik
% ...
\collaboration{XENON Collaboration}\email[]{xenon@lngs.infn.it}\noaffiliation

%% file: values.tex
\makeatletter
\newcommand{\setvalue}[2]{%
  \expandafter\def\csname value@#1\endcsname{#2}%
}

\newcommand{\get}[1]{%
  \csname value@#1\endcsname
}
\makeatother

\setvalue{nEventsSR0bkg}{7708}
\setvalue{nEventsSR1bkg}{5542}
\setvalue{nEventsSR1rn}{37835}
\setvalue{nEventsBetaTotal}{56965}
\setvalue{dTS1Cut}{25}
\setvalue{BetaEndPoint}{3.27}

%% file: sections/1_Introduction.tex
\section{\label{sec:introduction}Introduction}

\begin{figure}[tp]
  \centering
  \includegraphics[width=.9\linewidth]{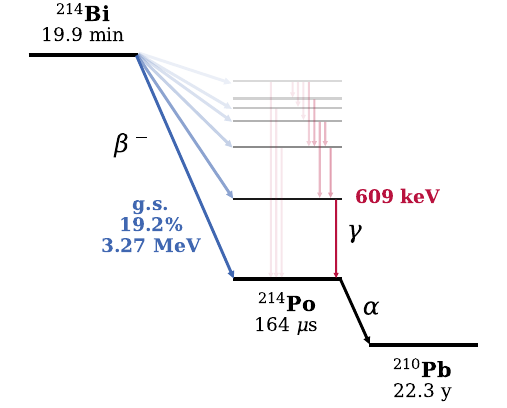}
  \caption{Decay scheme of \Bi\ to \Po\ and subsequently to \isotope[210]{Pb}. The diagram shows \bdecs\ from \Bi\ (left) either directly to the ground state (g.s.) or via excited states of \Po\ (center). Approximately 80 excited nuclear states are accessible during the decay (symbolically represented). All these excited states de-excite via \gdecs\ emissions (red arrows), always including at least one transition greater than \SI{609}{\keV}. The final step is the \adec\ of \Po\ (black arrow) directly to the stable ground state of \isotope[210]{Pb} (right).}
  \label{214BiDecayScheme}
\end{figure}

Rare-event searches require a precise characterization of backgrounds in the signal region of interest. An ubiquitous source arises from radon, which emanates from detector materials containing trace amounts of uranium and thorium. Radon progeny undergoing \bdec\ \cite{Feynman_Gell-Mann} constitute a major background, with \Pb\ and \Bi\ decays being key contributors in Weakly Interacting Massive Particle (WIMP) dark matter searches~\cite{XENONnT_SR1_WIMP, LZ, PandaX4T} and neutrinoless double beta-decay searches~\cite{XENONnT_highER, EXO-200, KamLAND-Zen}.

This study focuses on the \Bi\ \bdec\ spectral shape, whose decay scheme is shown in Fig.~\ref{214BiDecayScheme}. The isotope \Bi\ has a half-life of \SI{19.9}{\min}, a Q-value of \SI{\get{BetaEndPoint}}{\MeV}, and a \bdec\ branching ratio of $99.979\%$, of which $19.2\%$ decay directly into the ground state \cite{a214:WU2009681}. 

The Q-value at $\mathcal{O}$(MeV) promotes the transition into excited states. The subsequent de-excitation leads to complex gamma emission patterns. Theoretically, the multitude of accessible decay modes complicates the evaluation of the beta spectrum, as calculations must account for a broad spectrum of nuclear transitions. Understanding the nuclear structure of isotopes like \Bi\ is crucial across various fields. One example is the modeling of the rapid neutron-capture process (r-process) \cite{Arnould2007}, a key mechanism in stellar nucleosynthesis responsible for producing elements heavier than iron. 

Experimentally, having combined beta and gamma signals poses challenges in accurately measuring their individual contributions. However, a well-understood \Bi\ beta-spectrum would allow for a continuous detector calibration up to the beta-endpoint at \SI{\get{BetaEndPoint}}{\MeV}, which is not possible with any other calibration sources. 

In this study, we report a dedicated analysis in determining the \Bi\ ground-state spectrum using data from the XENONnT experiment \cite{XENONnT_2024}. Compared to the \Bi\ spectrum reconstruction by the NEMO experiment \cite{nemo214, nemo_bipo3}, the XENONnT detector has several advantages, including having an ultra-low background and an in-situ \Rn\ calibration. As the full \Rn\ decay process is contained within the XENONnT detector, the \adecs\ and \gdecs\ from the same decay chain are utilized to isolate the \Bi\ \bdecs. The selection algorithm developed in this work enables an effectively background-free measurement of the \Bi\ spectrum. Particular emphasis is placed on isolating the ground-state \Bi\ spectrum, allowing for direct comparisons with theoretical predictions. Furthermore, the XENON collaboration has developed a full-chain simulation framework, incorporating accurate detector response modeling. The simulation was adapted to develop the selection algorithm, while the \Bi\ data serves as a supporting benchmark for the simulation.

%% file: sections/2_Theory.tex
\section{\label{sec:theory}Theoretical Framework}

\Bdec s are interactions mediated by massive $W^{\pm}$ bosons between hadronic and leptonic currents. Assuming the decaying nucleon does not interact with the other $A-1$ nucleons, the probability density of the \bdec\ electron being emitted with kinetic energy $E_e$ can be expressed as \cite{Commins}:
\begin{align}
    P(E_e)dE_e &= \frac{G_\text{F}}{(\hbar c)^6}\frac{1}{2\pi^3\hbar}C(E_e)\notag\\
    &\times p_ecE_e(E_0-E_e)^2F_0(Z, E_e)dE_e,
\label{betaDistr}
\end{align}
where $p_e$ is the momentum of the electron, $E_0$ the \bdec\ endpoint energy, $Z$ the proton number, $G_F$ the Fermi constant, and $F_0(Z, E_e)$ the Fermi function. The nuclear structure information is encoded by the shape factor $C(E_e)$, which can generally be decomposed into vector, axial-vector, and mixed vector-axial-vector components:
\begin{align}
    C(\omega_e) &= g_V^2C_V(\omega_e) + g_A^2C_A(\omega_e) + g_Vg_AC_{VA}(\omega_e),
\label{shapeFactor}
\end{align}
where the unitless kinematic quantity $\omega_e = W_e/m_ec^2$ is used, and $g_V$ and $g_A$ are the corresponding coupling constants for the vector and axial-vector components \cite{Feynman_Gell-Mann}. The contribution of the shape factors $C_V$, $C_A$, and $C_{VA}$ is characterized by the nuclear matrix elements (NMEs).

The theoretical description of the \bdec\ has undergone significant advancements in recent years, which have been essential for the background modeling of the XENON experiments \cite{PhysRevC.102.065501, mougeot:cea-01809226}, addressing the challenges when moving towards heavier nuclei and higher \bdec\ Q-values. The complications include incorporating a large number of different transition modes in NMEs \cite{PhysRevC.109.014326}, modeling strong quenching of the weak coupling constant $g_A$ due to many-nucleon correlations \cite{Zhi2013} involving numerous approaches \cite{Deppisch2016, Kostensalo2017, Toivanen1998, Caurier2012, Iachello1991}, and applying the mesonic enhancement correction from the effects of nuclear medium \cite{Kostensalo2018}. 

\begin{figure}[tp]
  \centering
  \includegraphics[width=\linewidth]{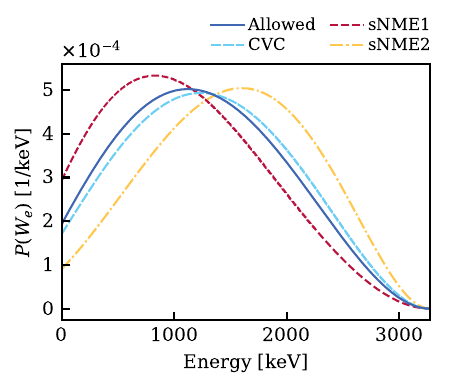}
  \caption{Theoretical \bdec\ energy spectra for ground-state to ground-state transitions of \Bi\ to \Po. The solid blue line shows the zeroth order \textit{Allowed} spectrum, the dashed light blue line the \textit{CVC}-based spectrum, the dashed red line the \textit{sNME1} fit, and the dot-dashed yellow line the \textit{sNME2} fit.}  
  \label{fig:models}
\end{figure}

For the \Bi\ \bdec, the transition to the ground state is classified as first-forbidden non-unique, which requires non-diagonal NMEs for its description~\cite{EJIRI20191}. The simplest reference is the zeroth-order spectrum, calculated analytically under the allowed approximation using the BetaShape code~\cite{mougeot:cea-01809226}, and shown in Fig.~\ref{fig:models} as \textit{Allowed}. A more complete theoretical description combines nuclear shell-model calculations with next-to-leading-order corrections under the conserved vector current (CVC) hypothesis. This approach enforces isospin symmetry of the strong interaction, fixes the weak vector coupling constant to $g_V=1.0$, and directly constrains the relevant vector form factors~\cite{Hardy2005,PhysRevC.95.024327,Kumar2021}. Building on the same framework, a recent study introduced a fit procedure in which the small relativistic vector NMEs (sNMEs) are varied simultaneously across all transitions to reproduce measured branching ratios~\cite{PhysRevC.109.014326}. From this study, we obtain the three additional spectra shown in Fig.~\ref{fig:models}: the \textit{CVC} prediction, and the two degenerate solutions of the sNME fit, labeled \textit{sNME1} and \textit{sNME2}. These results illustrate that the choice of sNME values can substantially modify the predicted spectral shape, making experimental input essential to discriminate between models.

%% file: sections/3_Experimental.tex
\section{\label{sec:experimental}Experimental Method}
\subsection{XENONnT Detector and Datasets}

The XENONnT experiment \cite{XENONnT_2024} uses xenon as both the target and detection medium inside a cylindrical dual-phase time projection chamber (TPC) that is \SI{1.33}{\meter} in diameter and \SI{1.49}{\meter} tall. The TPC contains \SI{5.9}{\tonne} of liquid xenon (LXe), with a thin layer of gaseous xenon (GXe) at the top. An interaction with xenon atoms inside the TPC produces both prompt scintillation light, defined as the S1 signal, and ionization electrons. Meshed electrodes are used to apply an electric drift field in LXe (\SI{23}{\volt/\centi\meter}) and an extraction field in the region of the liquid-gas interface (\SI{2.9}{\kilo\volt/\centi\meter}). This allows the free electrons to drift upward towards the LXe surface, where they are extracted into the GXe and generate secondary scintillation via electroluminescence, defined as the S2 signal. A total of 494 Hamamatsu R11410-21 photomultiplier tubes (PMTs) \cite{PMT_XENONnT:Antochi2021} are arranged at the top and bottom of the TPC to capture both the S1 and S2 signals, measured in units of photoelectrons (PE). A MeV beta-electron in the TPC generates an S1 with a signal area of approximately $10^4$ PE and an S2 of order $10^6$~PE, while a MeV alpha generates an S1 of several $10^4$~PE and an S2 of order $10^4$~PE.

An event consists of a pair of S1 and S2 signals. The pairing enables reconstruction of the deposited energy and the three-dimensional interaction position, where the xy-position is derived from the S2 scintillation photon hit pattern on the top PMT array and the z-position from the time difference between S1 and S2. The position reconstruction permits the definition of a fiducial volume, selecting only the data from the interior of the TPC, and thereby avoiding the background events originating from the decay of radioactive contaminants accumulated on the TPC wall. The dual signal also allows the discrimination between interactions with atomic electrons and with nuclei. The former is referred to as electronic recoil (ER) events, and the latter as nuclear recoils (NR) events. In dark matter searches, WIMP signals have the NR signature. In this work, the \Bi\ \bdec\ signal of interest has the ER signature.

To achieve a minimal background, the XENONnT experiment is operated at the INFN Laboratori Nazionali del Gran Sasso (LNGS) underground laboratory. The cryostat containing the TPC is the innermost of three nested detectors. It is enclosed by a neutron veto~\cite{NeutronVeto} that is itself nested within a muon veto. Both veto detectors are water Cherenkov detectors and are optically separated. All detector construction materials were selected for low radioactivity \cite{xenon2022_radiopurity}. The LXe is constantly purified via both gaseous and liquid purification systems \cite{XENONnT_2024} that remove electronegative impurities. The noble gas \KrEightFive\ \bdec\ isotope, which is intrinsically present in xenon, was removed via a dedicated cryogenic distillation system \cite{KrDistillation} with the remnant measured \cite{RGMS}. For the radon background, a radon removal system also based on cryogenic distillation was installed to continuously remove the radon remnant from the TPC \cite{radondistillation:XENON100:2017gsw}. Finally, degreasing, etching, passivating, and screening were performed to ensure low radon emanation and electronegative impurities for all TPC components \cite{xenon2022_radiopurity}. With all these background removal measures, XENONnT has achieved the lowest ever ER background among dark matter detectors \cite{radonRemoval}. 

This study includes 95.1 days of data from the first science run (SR0) and 186.5 days from the second (SR1) of the XENONnT experiment. These datasets contain \Bi\ decays from natural \Rn\ emanation from detector materials at a \Rn\ rate of approximately \SI{1.8}{\micro\becquerel/\kg} of LXe for SR0 and \SI{1.0}{\micro\becquerel/\kg} for SR1 \cite{radonRemoval_level}.

To obtain sufficient statistics for this study, the most relevant XENONnT dataset is from a calibration campaign during SR1 using a \Rn\ source (SR1 \isotope[222]{Rn}). The source was produced by implanting \Ra\ into a \SI{2}{\centi\meter}~$\times$~\SI{2}{\centi\meter} stainless steel plate at the ISOLDE facility at CERN \cite{radon222_source}, giving an overall emanated \Rn\ activity of about \SI{2}{\becquerel}. The implantation is sufficiently deep below the surface, such that the \Ra\ atoms are mechanically sealed within the sample. Only the noble gas decay daughters \Rn\ are ejected from the sample. 

\begin{figure}[tp]
  \centering
  \includegraphics[width=\linewidth]{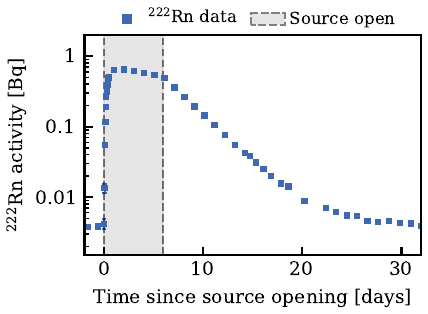}
  \caption{Time evolution of the measured $^{222}$Rn activity in the XENONnT TPC fiducial volume during the calibration campaign. The activity is derived from alpha-decays, shown as a function of time since the source opening. The vertical dashed lines indicate the source opening and closing times.}
  \label{fig:rn222calibration}
\end{figure}

During the \Rn\ (\Bi) calibration period, the stainless steel sample was installed in-line with the gaseous purification system \cite{XENONnT_2024} and was flushed with GXe to carry the \Rn\ atoms to the xenon TPC. The time evolution of the source activity is tracked in-situ via \Rn\ \adecs\ as shown in Fig.~\ref{fig:rn222calibration}. Shortly after the source injection, the \Rn\ activity increased by two orders of magnitude. Upon closing the source, the $^{222}$Rn level decreased by a compound effect of the \Rn\ natural decay and the continuous operation of the radon removal system, returning the \Rn\ activity to the level observed before the calibration campaign. The expected level of long-term contamination, particularly the \bdec\ isotope \PbTwoOneZero\ with a half-life of \SI{22.3}{yrs} that attaches to the TPC surfaces, has been evaluated as negligible.

We analyze the three datasets (SR0, SR1, SR1 $^{222}$Rn) with independent treatments to account for their differences in operation and calibration conditions, before combining them into the final spectral data.

Finally, to accurately estimate the ER yield model in-situ, we also utilized the XENONnT data from \isotope[220]Rn (\isotope[212]Pb) calibrations~\cite{radon220_calibration}. 

\subsection{Event Topology}

\begin{figure*}
  \centering
  \includegraphics[width=\linewidth]{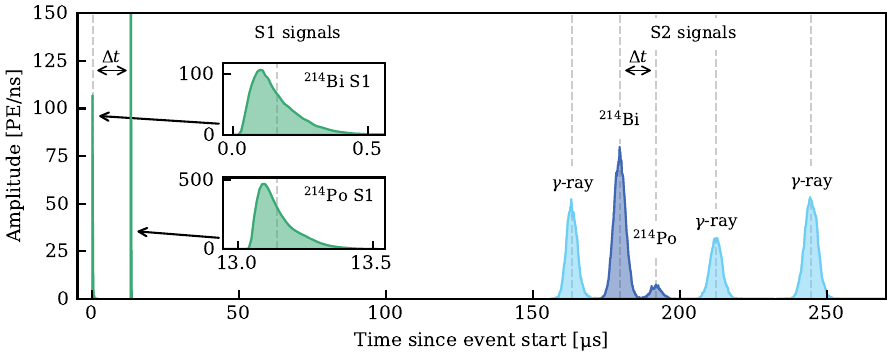}
  \caption{Waveform representation of a typical \BiPo\ event from measured data. The two green peaks on the left correspond to the S1 signals of \Bi\ and \Po, separated by a time interval $\Delta t$. The two dark blue peaks on the right, also separated by $\Delta t$, are identified as the corresponding S2 signals. The light blue peaks represent S2 signals from gamma emissions. Due to variations in electron drift distance, depending on whether the gamma travels upward or downward, its S2 signal appears earlier or later than the \Bi\ S2. The \adec\ (\Po) exhibits significant recombination and quenching effects, resulting in a smaller S2 despite its higher energy compared to the preceding \bdec . Conversely, these effects enhance the alpha’s S1 signal. Inset panels show the S1 waveforms on an expanded time scale, highlighting their much shorter duration compared to S2 signals.}
  \label{fig:waveforms}
\end{figure*}

To select \Bi\ decays to the \Po\ ground state, we exploit the characteristic event topology. The delayed coincidence of a \Bi\ \bdec, followed by the \Po\ \adec\ within the same chain is defined as \isotope[214]{BiPo}. Tagging an isolated \Bi\ \bdec\ is inaccessible because of its broad spectrum and the presence of other ER background components. In contrast, the \SI{7.83}{\MeV} \Po\ \adec\ \cite{a214:WU2009681} produces a prominent signal, with S1 signals of order $10^4$~PE. The large S1 arises from the high stopping power of alphas, which create dense electron-ion tracks where recombination suppresses the S2 but enhances the S1 \cite{AlphaInLXe}. 
This tagging allows the determination of the alpha rate by isolating and counting the alpha events with negligible background. The derivation of the \Rn\ activity in Fig.~\ref{fig:rn222calibration}, for instance, directly benefits from this tagging scheme. For the \Po\ \adecs, they are especially well-separated from the nearby alpha peaks by approximately \SI{1}{\MeV}. Moreover, the half-life of \Po\ is \SI{164}{\micro\second} \cite{a214:WU2009681}, which is relatively short compared to the full TPC drift time of $\sim$\,\SI{2}{\milli\second} and longer than the typical S2 signal width of $\sim$\,\SI{20}{\micro\second}. In other words, we can use the \Po\ alpha to tag the \Bi\ decay in a \BiPo\ event with effectively no background.

The \BiPo\ event topology is illustrated in Fig.~\ref{fig:waveforms}. The event is first triggered by the \Bi\ S1 signal and is shortly followed by a significantly larger S1 signal from \Po\ after a time difference of $\Delta t$. The S2s arrive a drift time later. The \Po\ S2 signal is suppressed as expected for an alpha event. The $\Delta t$ is effectively preserved by the S2s because the alpha signals are localized in position: the \Po\ alphas travel on an order of \SI{10}{\micro\meter} in LXe from the preceding \Bi\ decay. 

A complication arises when considering the \Bi\ \bdec\ into excited states that subsequently produce gammas from nuclear de-excitation. Given that the de-excitation half-life is in the nanosecond range, the S1s of these gammas are always merged with the S1 of the beta. However, unlike alpha particles, the gamma-rays can travel a few centimeters before interacting with another xenon atom~\cite{148746}. This shift in positions of the interaction site results in \BiPo\ events with more than two S2s, as illustrated in Fig.~\ref{fig:waveforms}. Notably, however, among all the de-excitation gammas of the \Bi\ \bdec\ as illustrated in Fig.~\ref{214BiDecayScheme}, there is always one gamma photon with an energy at least \SI{609}{\keV}. We can effectively use these high-energy \gdecs\ to tag the excited states.

\subsection{Event Selection}
\label{subsec:BiPoAlg}

\par This analysis relies on the \textit{\BiPo\ tagging algorithm}, a method that exploits the time coincidence between the \Bi\ \bdec\ and the subsequent \Po\ \adec. Although \BiPo\ tagging is commonly employed in low-background experiments~\cite{PhysRevD.110.012011, PhysRevLett.117.082503, PhysRevD.92.072011}, our implementation specifically addresses the complexity introduced by gamma emissions accompanying \Bi\ decays into excited nuclear states. These emissions produce events with multiple signals, complicating straightforward time-based matching. To reliably pair the correct S1 and S2 signals, our algorithm compares the time differences $\Delta t$ between all combinatoric pairs formed by the three largest S1 signals (to account for possible spurious signals) and up to ten largest S2 signals. We identify events in which one S1 pair and one S2 pair have matching $\Delta t$ values within a tolerance of $\pm$\SI{3}{\micro\second}.
Peaks from S2 signals with areas below \SI{1500}{PE}, unphysical top-bottom PMT area distributions, or inappropriate timing (such as spurious pulses that occur shortly after the main pulse and can be reconstructed as separate peaks), are excluded from the matching procedure. We use simulations, described in Sec.~\ref{sec:simulation}, to quantify the matching efficiency, which is $>91\%$ for true ground-state events, providing validation of the matching approach.

Following the matching procedure, we apply a series of \textit{quality selection criteria} to ensure high-quality event reconstruction and minimize contributions from backgrounds or poorly reconstructed signals. Events must pass the fiducial volume constraint, defined by a vertical depth at least \SI{5}{cm} away from the top and bottom boundaries of the TPC to avoid regions with field inhomogeneities~\cite{FieldCage}. A minimum separation of \SI{\get{dTS1Cut}}{\micro\second} between the two primary S1 signals is required to prevent potential merging of the corresponding S2 signals, given a typical S2 width at $90\%$ of the area of up to \SI{20}{\micro\second}. This criterion results in only a modest loss of approximately $3\%$ of events, as shorter S1 intervals typically produce merged, unresolved S2 signals. This requirement also eliminates the background contribution of \isotope[212]{BiPo} events originating from the \isotope[220]Rn decay chain, with a similar topology as the \BiPo\ but with a half-life of \SI{300}{\nano\second}. Additional criteria based on the alpha S1 signal area are employed to remove events originating from other \adecs. These selection steps reduce the dataset to approximately \SI{165000} events, which include both ground-state and excited-state decays.

\par The final \textit{ground-state selection} step specifically isolates \Bi\ decays directly to the \Po\ ground state. We require exactly two S2 signals above \SI{1500}{PE}, corresponding to the \bdec\ and the subsequent \adec. To suppress events with additional gamma interactions, any third S2 is required to be smaller than \SI{10000}{PE}. This selection effectively excludes most excited-state decays, which, due to selection rules, de-excite through high energy \gdecs\, leading to clearly identifiable additional S2 signals. Although further criteria, such as a selection on the S2 width to remove partially merged signals, were tested, they provided no significant improvement beyond the initial matching procedure. This selection yields \SI{\get{nEventsBetaTotal}} events, corresponding to 31\% of the events that passed the quality criteria.

\par Using detailed simulations (described in Sec.~\ref{sec:simulation}), we evaluate the performance and purity of the selection procedure. After applying matching and quality criteria, the requirement of precisely two S2 signals achieves a signal acceptance of around $94\%$ (the fraction of true ground-state events passing the final selection) and a background rejection of approximately $93\%$ (the fraction of excited-state events removed by the selection).
\par The fraction of excited-state events that pass the ground-state selection criteria (approximately $20\%$ of the total selected events) predominantly belongs to two categories of roughly equal proportions. The first category consists of events in which the emitted gammas completely escape the active detection volume. These events are reconstructed at energies lower than genuine ground-state events, as only the energy deposited by the beta electron is observed. The second category comprises events where gamma-induced signals fully merge with the beta electron signals, resulting in artificially higher reconstructed energies due to the combined gamma and beta contributions. Both categories distort the measured shape of the ground-state \Bi\ decay energy spectrum. As we compare simulated \Bi\ \bdec\ spectra to measured data, we model these leakage events from decays to excited states in addition to the ground-state decays.

\subsection{Simulations and Signal Modeling}
\label{sec:simulation}

Simulations for this analysis are produced using the XENONnT framework~\cite{fuse}, which incorporates a full-chain simulation from energy deposition to final event reconstruction. Initial interactions, including both the \Bi\ and subsequent \Po\ decays, are simulated with Geant4~\cite{GEANT4:2002zbu,1610988}. This step considers all relevant nuclear decay characteristics, including the full decay scheme of \Bi, associated branching ratios, and emitted gamma energies. To investigate different theoretical models of the \Bi\ ground-state \bdec~ spectrum introduced in Fig.~\ref{fig:models}, we adjusted the input beta energy distribution provided to Geant4 accordingly, creating simulation datasets for each considered model.

\par Subsequent stages of the simulation that include clustering of deposited energies, generation and propagation of photons and electrons using detector-specific efficiency maps, electron drifting, and waveform simulation are performed using the fuse package~\cite{fuse}. These simulated waveforms closely replicate those recorded by the data acquisition system of the XENONnT detector, allowing for the use of an identical processing pipeline as for the real data with the straxen package~\cite{straxen}.

\par For accurate modeling of events with energies up to \SI{\get{BetaEndPoint}}{\MeV}, which is substantially higher than the typical range targeted by the dark matter search analyses ($<$\SI{100}{keV})~\cite{PhysRevD.111.062006}, we extended the existing fuse framework with a specialized simulation workflow. Two critical improvements were implemented: a more physically motivated clustering algorithm for energy depositions, and a data-driven yield model specifically tailored for high-energy electron recoils. The latter describes the photon and electron yields per unit energy.

\par Clustering of energy depositions is necessary because the data-driven yield models, such as the ones provided by NEST~\cite{nest:model1, nest:review}, require input energies to attribute the correct amount of photons and electrons generated by individual interactions. Due to the incomplete understanding of the underlying microphysics at these scales, energy depositions must be aggregated into physically meaningful clusters before applying the yield model. The default clustering method used in XENONnT~\cite{PhysRevD.111.062006} groups energy depositions solely based on spatial and temporal proximity. While adequate at lower energies, this method is inadequate in describing complex high-energy event topologies. The new framework employs a lineage clustering method inspired by NEST, which groups interactions based on their physical properties and interaction history. Energy depositions from electron tracks are grouped into single clusters, except for Bremsstrahlung photons, which form separate clusters. Gamma interactions instead initiate new clusters for each distinct interaction site, effectively differentiating gamma from beta events.

\begin{figure}[tph!]
  \centering
  \includegraphics[width=\linewidth]{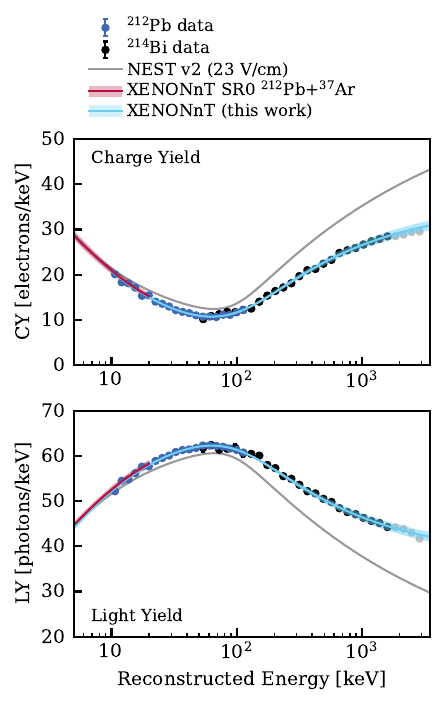}
    \caption{Charge yield (CY, top) and light yield (LY, bottom) as a function of reconstructed energy, extracted from \bdecs\ of \isotope[212]{Pb} (blue) and \Bi\ (black). The ER yield model (light blue) is obtained by fitting the charge yield, with the light yield derived as the complementary fraction of total quanta. The shaded band accounts for the systematic uncertainty from the energy-bias correction. A correction for residual excited-state leakage is applied to the data. Grayed-out \Bi\ points above \SI{1.8}{MeV} are excluded from the fit due to gamma contamination. For comparison, the default NEST model at \SI{23}{V/cm} (gray) and the XENONnT low-energy yield model (red) are shown. Statistical uncertainties are smaller than the marker size.}
  \label{fig:yields}
\end{figure}

\par Unlike the default NEST model, which employs separate yield models for gamma and beta interactions, our approach uses a single, unified yield model. The apparent difference between the two cases originates from the underlying physics: gamma-rays can interact via Compton scattering and therefore deposit their energy at multiple spatially separated sites, while electrons typically deposit all energy in a contained interaction cluster. Each site contributes its own cluster with energy $E_i$, and since the charge and light yields depend non-linearly on $E_i$, the combined yield of these clusters differs from that of a single deposition with the same total energy. In our analysis, this distinction emerges naturally from clustering: the yield difference between \bdecs\ and \gdecs\ is not imposed through separate models, but arises directly from the physical interaction topology when combined with the clustering method.

\par We extract a new parametrisation of the ER yield model using single \bdec\ events, which provide well-defined single-cluster energy depositions. For this purpose, we use \isotope[212]{Pb} decays from \isotope[220]{Rn} calibrations together with SR0 \Bi\ decays from this analysis, ensuring broad coverage across the full energy range. The yield parameters are obtained by fitting the ten-parameter NEST function~\cite{nest:model1} to the measured charge yield, with the light yield calculated as the complementary fraction of quanta. 

\par We account for residual leakage of excited-state events into the ground-state selection by comparing yields in simulation from truth-level ground-state \bdec\ events with those obtained when the excited-state events that passed the selection are included. The observed difference is implemented as an effective correction to the charge and light yield in data. Since the leakage events concentrate above \SI{1.8}{\MeV}, this region is excluded from the fit.

\par Figure~\ref{fig:yields} compares the resulting charge and light yield curves with both the default NEST model at a drift field of \SI{23}{V/cm} and the official XENONnT yield model used for low-energy searches \cite{PhysRevD.111.062006}. The latter, derived from \Ar\ and \isotope[212]{Pb} calibrations, is in good agreement with our parametrisation below \SI{20}{\keV}. The NEST model, although nominally valid at the MeV scale, is largely based on data below \SI{100}{\keV}. Moreover, only limited experimental input exists at the drift field of XENONnT (\SI{23}{V/cm}), making the extrapolation in this regime uncertain and leading to clear discrepancies in charge and light yields.

\par To account for the fact that the energy bias correction described in the next section acts on the reconstructed energy but not on S1 and S2 independently, we include in Fig.~\ref{fig:yields} a systematic uncertainty band on our yield curves. The shaded band reflects the propagation of the energy bias into the extracted yields, covering the systematic uncertainty introduced by peak reconstruction and subsequent energy determination.

\par We validated the resulting yield model using two independent datasets: the clean single-cluster beta selection described above, and the \BiPo\ events including excited-state decays that produce combined beta and gamma topologies. These events, with multiple interaction sites and complex clustering, provide a stringent test of the clustering and yield framework. Although the yield parameters were derived only from clean \bdecs\ below \SI{1.8}{\MeV}, the model reproduces the measured S1 and S2 distributions at the percent level across the full energy range, up to the \Bi\ endpoint. By contrast, simulations based on the default NEST yields and spatial clustering showed discrepancies of up to $50\%$ in S2 for beta–gamma events above a few hundred keV. The agreement obtained here therefore supports the changes in both the clustering method and the yield parametrisation: reproducing multi-cluster topologies at high energy is only possible if both components are accurate.

\par The clustering procedure and yield parametrisation introduced above are designed specifically to achieve accurate reproduction of the individual S1 and S2 distributions between data and simulations. They do not directly guarantee agreement in the reconstructed energy spectra, which is instead addressed in the following section through dedicated corrections.

\subsection{Energy Bias Correction}
\label{sec:bias}

To ensure consistent energy reconstruction across simulations and the three experimental datasets (SR0, SR1, and SR1 \isotope[222]{Rn}), we derive dataset-specific energy-dependent bias corrections, addressing known effects described in previous XENONnT analyses~\cite{PhysRevD.111.062006}. These biases affect both the S1 and S2 signals, primarily arising from electronic noise, PMT response effects, afterpulses, photoionization, and threshold effects from the data acquisition system and ancillary software thresholds.

\begin{figure}[t!]
\centering
\includegraphics[width=\linewidth]{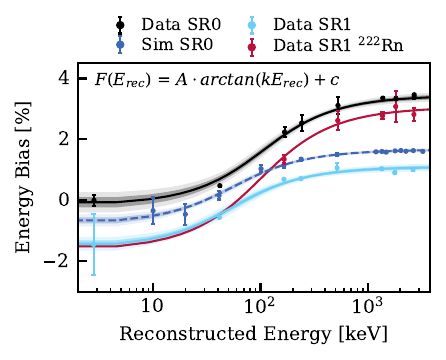}
\caption{Energy bias as a function of reconstructed energy for SR0 (black), SR1 (cyan), and SR1 \isotope[222]{Rn} calibration data (red), as well as SR0 simulations (blue dashed). The data points correspond to monoenergetic calibration peaks at known energies. The uncertainties on each point correspond to the uncertainty in the mean obtained from the Gaussian fit to the energy distribution of the selected peaks.  Bias values are fitted with an arctangent function, and shaded bands indicate the 1$\sigma$ and 2$\sigma$ confidence intervals of the fit. Uncertainty bands for the \isotope[222]{Rn} dataset are omitted for clarity.}
\label{fig:energy-bias}
\end{figure}

\par To accurately compare true energies, we first reconstruct the energy for each event as:
\begin{equation}
E_{\text{rec}} = W \cdot \left( \frac{cS1}{g_1} + \frac{cS2}{g_2}\right),
\end{equation}
where $W =$~\SI{13.7}{\eV}~\cite{S1S2signal} is the average energy required to produce one quantum (photon or electron) in liquid xenon, cS1 and cS2 are the corrected S1 and S2 signals, and $g_1$ and $g_2$ are detector-specific calibration constants determined from the data~\cite{PhysRevD.111.062006}. In this analysis, we use the following values that were determined in earlier XENONnT analyses. For SR0 we adopt $g_1 = 0.151$ PE/photon and $g_2 = 16.5$ PE/electron~\cite{PhysRevD.111.062006}; for SR1, we use $g_1 = 0.1367$ PE/photon and $g_2 = 16.9$ PE/electron~\cite{XENONnT_SR1_WIMP}.

\par Following the procedure described in \cite{PhysRevD.111.062006}, we then define an empirical energy correction for each dataset by measuring the bias using monoenergetic peaks from background and calibration sources. For each peak, the bias is calculated using $\Delta E = (E_{\text{rec}}-E_{\text{true}})$, and fitted with an arctan function:
\begin{equation}\label{biasfunc}
F(E_{rec}) \equiv \frac{\Delta E}{E_{\text{rec}}} \times 100 = A \cdot \arctan(k E_{\text{rec}}) + c,
\end{equation}
where $A$, $k$, and $c$ are free fit parameters. 

\par We use monoenergetic lines from \isotope[37]{Ar} (\SI{2.8}{\keV}), \isotope[83\text{m}]{Kr} (\SI{41.5}{\keV}), \isotope[131\text{m}]{Xe} (\SI{163.9}{\keV}), \isotope[129\text{m}]{Xe} (\SI{236.1}{\keV}), electron-positron annihilation (\SI{511}{\keV}), \isotope[60]{Co} (\SI{1173.2}{\keV} and \SI{1332.5}{\keV}), gamma-rays from external\footnote{External \Bi\ refers to the \Bi\ produced in the uranium decay chain within the construction materials. In this case, only the gamma-ray can reach the TPC, while the accompanying electron is absorbed before entering the detector and remains undetected.} \Bi\ (\SI{1764.5}{\keV}), and \isotope[208]{Tl} (\SI{2614.5}{\keV}). For SR1, the \isotope[37]{Ar} line originates from an accidental injection of commercial-grade xenon into the detector before the start of SR1. The \isotope[37]{Ar} was removed using cryogenic distillation before the \isotope[222]{Rn} calibration, so no \isotope[37]{Ar} datapoint is available in the SR1 \isotope[222]{Rn} dataset. For simulations, monoenergetic gamma-rays and electrons at different energies were uniformly simulated throughout the detector volume.

\par Figure~\ref{fig:energy-bias} shows the fitted bias curves for each dataset, exhibiting clear differences: SR0 is calibrated such that \isotope[37]{Ar} has zero bias by construction, while SR1 shows systematically lower biases, becoming negative at low energies. The SR1 \isotope[222]{Rn} calibration data display higher biases at higher energies. We attribute the overall offset of the curves mainly to different choices in the $g_1$ and $g_2$ determination procedure, while the increase in amplitude with energy is governed by detector conditions such as the higher event rate during calibrations and photoionization, where absorbed photons release additional delayed electrons. Because no \isotope[37]{Ar} anchor point is available in the SR1 \isotope[222]{Rn} dataset, the fit was performed under constraints to prevent unphysical divergence. For this reason, we do not show the uncertainty band for this dataset in Fig.~\ref{fig:energy-bias}, as the unconstrained errors would be unrealistically large. This choice does not affect the analysis: the lowest energy bin used in the spectrum study is at \SI{65}{\keV}, well above the missing \isotope[37]{Ar} line. 

\par The best-fit parameters for each dataset are reported in Tab.~\ref{tab:biasres} in the appendix.

%% file: sections/4_Results.tex
\begin{figure*}[tph!]
  \centering
    \adjustbox{valign=t}{\includegraphics[width=.49\linewidth]{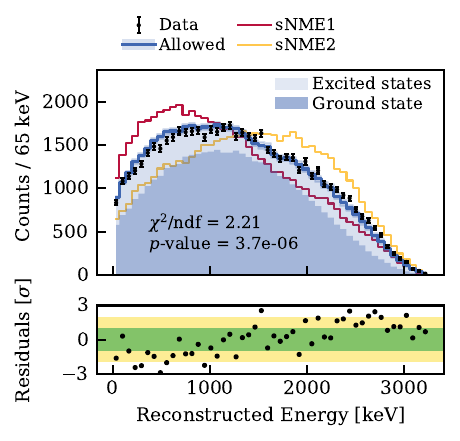}}
    \adjustbox{valign=t}{\includegraphics[width=.49\linewidth]{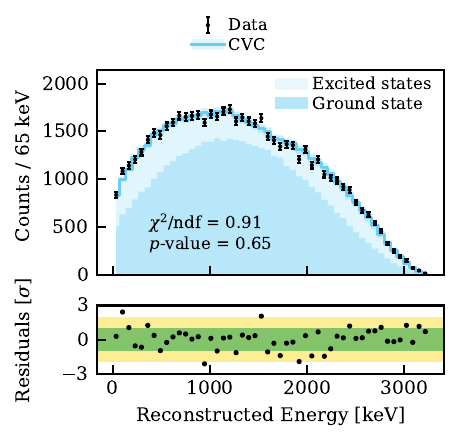}}
  \caption{Detailed comparison of the measured energy spectrum of \Bi\ ground-state data with simulations. 
  Left panel: Data compared with the allowed model (filled histogram), including the true ground-state and excited-state contributions. The sNME1 and sNME2 theoretical spectra are shown as thin lines for reference. Right panel: Comparison with the CVC spectrum (filled histogram). The normalised residuals with respect to the allowed (left) and CVC (right) spectra are displayed in the bottom panels, along with the corresponding reduced chi-squared ($\chi^2/\text{ndf}$) and p-values.} 
  \label{fig:results}
\end{figure*}

\section{\label{sec:results}Results}

\subsection{Models to Data Comparison}

In this analysis, the event selection is rendered effectively free of background contamination as a result of the \BiPo\ identification algorithm. The primary remaining background arises from gamma leakage from excited-state decays into the ground-state selection. Since this leakage is accurately modeled by our simulations, which incorporate well-measured gamma emissions from \Bi, we can compare data directly to full-chain simulations.

We combine data from the three experimental datasets: SR0 (\SI{\get{nEventsSR0bkg}} events), SR1 (\SI{\get{nEventsSR1bkg}} events), and the SR1 \Rn\ (\SI{\get{nEventsSR1rn}} events). The consistency of the ratio between quality-selected and ground-state-selected events across these datasets ($\sim30\%$) allows combining them into a single dataset. The energies of each dataset have been individually corrected for the energy bias described in Sec.~\ref{sec:bias}.

\par Figure~\ref{fig:results} presents the combined dataset in comparison with four simulation sets corresponding to the different theoretical \bdec\ spectra: \textit{Allowed}, \textit{CVC}, \textit{sNME1}, and \textit{sNME2}. The simulated spectra are normalized to match the total number of events observed in data; this normalization involves no free fitting parameters, ensuring a robust, unbiased comparison of spectral shapes. Statistical and systematic uncertainties are included in the simulations, as discussed in Sec.~\ref{sec:systematics}, whereas the data points represent purely statistical uncertainties.

\par The sNME1 and sNME2 models exhibit significant discrepancies from the measured spectrum and are excluded from further consideration. The allowed model shows a clear energy-dependent deviation visible in the residuals (bottom panel), indicating a systematic mismatch. On the other hand, the CVC spectrum provides an excellent description of the data, with a high p-value larger than $0.05$ and no evident residual energy trend. This comparison strongly favors the CVC spectrum as the most accurate theoretical description of our data.

\subsection{Systematic Uncertainties}\label{sec:systematics}

We address potential systematic uncertainties arising from mismatches between simulation and data, particularly those with energy-dependent effects. We identify and evaluate three primary sources of uncertainty.

\par First, we consider uncertainties due to gamma-event leakage into the beta selection. While waveform properties like the S1/S2 amplitudes and signal widths are well matched between simulations and data, slight mismatches could still affect the fraction of excited-state events leaking into the ground-state selection. To account for this, we apply a $1\%$ variation to the total number of leakage events. This choice is motivated by the observed stability (within 1\%) of the ratio of selected ground-state events to all quality-selected events across multiple datasets.

\par Second, we examine uncertainties related to the energy scale of excited-state events. While gamma energies are well known from precise nuclear spectroscopy, the associated beta spectra are more difficult to measure and model. To assess the stability of the result, we apply an effective $5\%$ energy shift to all excited-state events. This corresponds to approximately $150$\,keV at the beta endpoint, compatible with the spectral shifts observed in the ground-state spectrum studied in this work. The effect on the reconstructed spectrum is limited by the relatively small contribution of excited-state events and the dominant role of the well-measured gamma energies.

\par Finally, we evaluate the overall energy scale uncertainty. The reconstructed energy is corrected using energy-bias curves derived from calibration peaks, for which the fitted peak positions have uncertainties below $0.25\%$. To account for possible residual effects, we apply a conservative $0.5\%$ global energy shift. At the beta endpoint, this corresponds to a less than \SI{20}{\keV} shift, which is smaller than the bin widths.

\par These three sources of uncertainty were combined by evaluating their impact on the simulated energy spectrum in each energy bin. The resulting variations were added in quadrature and incorporated as an additional uncertainty on the simulated prediction, on top of the statistical uncertainties from data. The combined systematic uncertainty remains below $1\%$ across the entire energy range and is subdominant to the statistical error. In the high-energy tail, where event statistics are low, systematic and statistical effects become correlated, but statistical uncertainties dominate.

%% file: sections/5_Conclusion.tex
\section{\label{sec:conclusion}Conclusion and outlook}
We presented a measurement of the energy spectrum of the \Bi~ beta-decay to the ground state of \Po~ using data from the XENONnT experiment. The event selection employs a dedicated \BiPo\ tagging technique, combining background data with a \Rn\ calibration campaign. By developing an improved high-energy simulation pipeline, we successfully modeled ER signals up to \SI{\get{BetaEndPoint}}{\MeV}, well beyond the low-energy range defined for WIMP searches. Our analysis strongly favors the theoretical spectrum calculated under the conserved vector current (CVC) hypothesis, significantly rejecting the allowed approximation and alternative nuclear models. These results demonstrate the capability of dual-phase xenon detectors to perform high-precision beta spectroscopy, providing valuable validations for theoretical predictions and reinforcing their potential for probing nuclear processes relevant to fundamental physics and astrophysics.

%% file: sections/9_Appendix.tex
\appendix
\section{Yields Parameters}

\begin{table*}[th!]
\centering
\caption{Fit results for the NEST ER yield model parameters.}
\label{tab:nest_fit_results}
\begin{tabular}{clccc}
\toprule
\textbf{Parameter} & \textbf{Description} & \textbf{Fit Value} & \textbf{NEST Default (23 V/cm)} \\
\midrule

$m_1$ & Stitching-region yield & $5.00 \pm 0.24$ & $7.10$ \\
$m_2$ & Low-energy asymptote & $83 \pm 3$ & $77.3$ \\
$m_3$ & Thomas-Imel shape control & $2.00 \pm 0.06$ & $0.716$ \\
$m_4$ & Low-energy shape control & $0.77 \pm 0.03$ & $1.83$ \\
$m_5$ & High-energy asymptote & $29.3 \pm 0.7$ & $48.6$ \\
$m_7$ & High-energy scaling & $125.4 \pm 2.2$ & $94.4$ \\
$m_8$ & High-energy shape control & $2.21 \pm 0.08$ & $4.29$ \\
$m_9$ & Low-energy asymmetry control & $1.072 \pm 0.006$ & $0.334$ \\
$m_{10}$ & High-energy asymmetry control & $0.29 \pm 0.09$ & $0.066$ \\
\bottomrule
\end{tabular}
\end{table*}

\begin{table*}[th!]
    \caption{Energy Bias Fit Results}
    \label{tab:biasres}
\centering
\begin{tabular}{cccc}
\toprule
\textbf{Dataset} & \textbf{A} & \textbf{k} & \textbf{c} \\ \hline
SR0 & 2.30 ± 0.12 & 0.0095 ± 0.0013 & -0.2 ± 0.2 \\
SR1 & 1.70 ± 0.09 & 0.018 ± 0.003 & -1.56 ± 0.13 \\
SR1 $^{{222}}$Rn & 3 ± 7 & 0.01 ± 0.03 & -2 ± 12 \\
Simulations & 1.56 ± 0.07 & 0.018 ± 0.002 & -0.79 ± 0.12 \\
\bottomrule
\end{tabular}
\end{table*}

We summarize the best-fit parameters for the ER yield model obtained from fitting \isotope[212]{Pb} and \Bi\ calibration data in XENONnT. The model defines the charge yield as: 

\begin{multline}
Q_y(E, \mathcal{E})
=
m_1(\mathcal{E})
+
\frac
{m_2-m_1(\mathcal{E})}
{\left[1+\left(\frac{E}{m_3(\mathcal{E})}\right)^{m_4(\mathcal{E})}\right]^{m_9}}
\\ 
+
m_5(\mathcal{E}) 
-\frac
{m_5(\mathcal{E})}
{\left[
    1+
    \left( \frac{E}{m_7(\mathcal{E})}\right)
    ^{m_8}
    \right]
    ^{m_{10}(\mathcal{E})
}
}.
\end{multline}
where $E$ is the energy of the interaction and $\mathcal{E}$ is the electric field. The parameters $m_1$–$m_{10}$ are described in detail in Ref.~\cite{nest:review}.

Our best-fit values, reported in Tab.~\ref{tab:nest_fit_results}, differ notably from the default NEST values, particularly in the high-energy region ($m_5$, $m_8$, $m_{10}$), highlighting the need for a data-driven recalibration for accurate high-energy ER modeling in XENONnT.

\section{Energy Bias Fit Results}

In Tab.~\ref{tab:biasres} we report the parameters obtained from the fits of the energy bias function described in Sec.~\ref{sec:bias}. The function and the parameters are introduced in Eq.~\ref{biasfunc}. The reported best-fit values can be used to reproduce the bias correction curves shown in Fig.~\ref{fig:energy-bias} for SR0, SR1, SR1 \isotope[222]{Rn}, and the SR0 simulation dataset.